\documentclass[aps,showpacs,preprintnumbers,amsmath,amssymb, 
twocolumn, tightenlines,
]{revtex4}

\usepackage[dvips]{graphicx}\usepackage{bm}
\usepackage{amsmath}
\usepackage{dcolumn}
\usepackage{bm}

\begin{document}

\bibliographystyle{apsrev} 

\title {When renormalizability is not 
  sufficient: Coulomb problem for vector bosons}

\author{V.V.Flambaum} \email[Email:]{flambaum@phys.unsw.edu.au}
\author{M.Yu.Kuchiev} \email[Email:]{kuchiev@phys.unsw.edu.au}

\affiliation{School of Physics, University of New South Wales, Sydney
  2052, Australia} \affiliation{Physics Division, Argonne National
  Laboratory, Argonne, Illinois 60439-4843, USA}

    \date{\today}

    \begin{abstract} The Coulomb problem for vector bosons
      $W^\pm$ incorporates a known difficulty; the boson falls on the
      center.  In QED the fermion vacuum polarization produces a
      barrier at small distances which solves the problem. In a
      renormalizable $SU(2)$ theory containing vector triplet
      $(W^+,W^-,\gamma)$ and a heavy fermion doublet $F$ with mass $M$
      the $W^-$ falls on $F^+$, to distances $r \sim 1/M$, where $M$
      can be made arbitrary large.  To prevent the collapse the theory
      needs additional light fermions, which switch the ultraviolet
      behavior of the theory from the asymptotic freedom to the Landau
      pole. Similar situation can take place in the Standard Model.
      Thus, the renormalizability of a theory is not sufficient to
      guarantee a reasonable behavior at small distances for
      non-perturbative problems, such as a bound state problem.
      \end{abstract}
    
    \pacs{12.15.Ji, 12.15.Lk, 12.20.Ds}

    \maketitle

    It is usually believed that a renormalizable theory automatically
    exhibits good physical behavior at large momenta and small
    distances.  This claim is definitely correct in low orders of the
    perturbation theory. However, in higher orders, which are
    necessary, for example, in a bound state problem, the situation is
    not so obvious. We present here an example, in which the
    renormalizability by itself fails to define a proper behavior of
    the theory at small distances.
    
    Consider a negatively charged vector boson, which propagates in
    the Coulomb field created by a heavy point-like charge $Z|e|$
    assuming that the boson is massive. A bound state problem for this
    boson needs summation in all orders in $Z \alpha$.  Since the
    electrodynamics for massive vector particles is
    non-renormalizable, one should expect problems here. One of them,
    found long time ago, is particularly interesting for our
    discussion.  Soon after Proca formulated theory for vector
    particles \cite{proca_1936} it became clear that it produces
    inadequate results for the Coulomb problem: the $W$ wave function
    is so singular that the integral over the charge density of $W$ is
    divergent near the origin
    \cite{massey-corben_1939,oppenheimer-snyder-serber_1940,tamm_1940-1-2}.
    Corben and Schwinger \cite{corben-schwinger_1940} modified the
    Proca theory, tuning the Lagrangian and equations of motion for
    vector bosons in such a way as to force the gyromagnetic ratio of
    the vector boson to acquire a favorable value $g=2$. It is well
    known now that $g=2$ is the gyromagnetic ratio of the $W$-boson in
    the Standard model. This modification allowed Corben and Schwinger
    to obtain a physically acceptable spectrum for the Coulomb
    problem, which is described by the Sommerfeld formula similar to
    the spectrum of Dirac particles, but with integer values of the
    total angular momentum $j=0,1,...$.
  
    Corben and Schwinger found also that their modification did not
    resolve the main problem, the $W$-boson still falls to the center
    for two series of quantum states; one with $j=0$, and the other
    one with ``$l$''$=0$ (if ``$l$'' is defined appropriately). The
    wave function of $W$ is so singular at the origin for these states
    that the integral over $W$ charge density is divergent near the
    origin. In our works \cite{kuchiev-flambaum-06} we found a cure
    for this problem. The QED fermion vacuum polarization was shown to
    produce an effective potential barrier for the $W$ boson at small
    distances.  The charge density of the $W$ boson in the $j=0$ state
    decreases as $\exp{(-const/r)}$ under this barrier and vanishes at
    the origin; similar improvement exhibits the $l=0$ state. As a
    result the Coulomb problem for vector particles becomes well
    defined.  The corresponding correction to the Sommerfeld spectrum
    proves to be small.  The effective potential of Ref.
    \cite{kuchiev-flambaum-06} is repulsive only when the running
    coupling constant exhibits the Landau-pole behavior; in contrast,
    for asymptotic freedom, the collapse is inevitable.
      
    In \cite{kuchiev-flambaum-06} we derived the Corben-Schwinger
    Lagrangian from the Lagrangian of the Standard Model, where the
    mass of $W$ is produced by the Higgs mechanism, which preserves
    the renormalizability of theory. However, the applicability of the
    Corben-Schwinger wave equation to $W$ requires that the Coulomb
    center does not interact with the $Z$-boson and Higgs particle.
    (It may be taken, for example, as a small charged black hole.)
    However, such Coulomb center is not described by the Standard
    Model, preventing the theory from being a complete renormalizable
    one.
    
\begin{figure}
\centering \includegraphics[height=6.5
   cm,keepaspectratio=true]{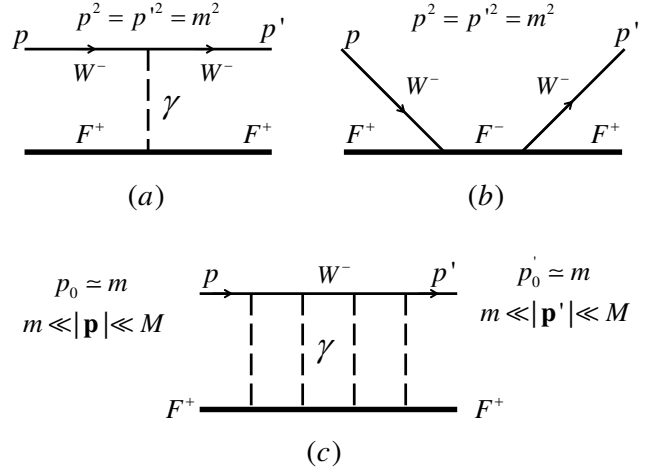}
   \caption{ 
      \label{one} 
      Scattering of the $W^-$-boson on the heavy fermion $F^+$, (a)
      the photon exchange, (b) - Compton-type scattering; (a) and (b)
      compensate each other for high energy of $W$ on the mass shell
      $p^2=m^2$; (c) one of the ladder-type diagrams, which are
      important for a short-distance behavior of the $W$-boson wave
      function at $ m^2\ll -p^2\ll M^2$.  }
\end{figure}

    In the present work we consider an example of a fully
    renormalizable model, which exhibits a similar phenomenon.  Take
    an $SU(2)$ gauge theory and a triplet of real Higgs scalars $\Phi$
\begin{eqnarray}
        \label{triplet} {\cal L }_\mathrm {Boson} =
-\frac{1}{4}\,G_{\mu\nu}^a\,G^{a\,\mu\nu}+ \frac{1}{2}\,D_\mu
\Phi^{a\,*} D^\mu\Phi^a+...~.  
   \end{eqnarray} 
   Here $G_{\mu\nu}^a$ and $D_\mu$ are the gauge field and the
   covariant derivative, which includes the gauge potential $A_\mu^a$;
   the dots refer to the nonlinear self-interaction of scalars, which
   produces the vacuum expectation value for the Higgs field
   $\Phi_\mathrm{vac} =(0,0,v)$.  Then one gauge boson $A_\mu^3$
   remains massless, call it the photon, $A_\mu=A_\mu^3$.  The two
   other bosons, which we call $W^{\pm}_\mu=(A^1_\mu \mp i
   A^2_\mu)/\sqrt{2}$ acquire the Higgs mass $m=g^2v^2/2$, see e.g.
   \cite{cheng}.  To allow the Coulomb center to appear in the model,
   consider a heavy fermion doublet $F=(F^+,F^-)$ with charges $e/2$
   and $-e/2$ for its two components.  Presume for simplicity that the
   parity is conserved and the fermion doublet does not interact with
   the Higgs field; its large mass, $M\gg m$, is a free parameter in
   the Lagrangian
\begin{eqnarray}
        \label{mass}
        {\cal L}_\mathrm{Fermi}= \bar{F}\,\left(\, i \gamma_\mu D^\mu-M \,\right)\,F~.
   \end{eqnarray}
   Our goal is to demonstrate that the interaction between $W^-$ and
   the heavy fermion $F^+$ results in the collapse of the boson onto
   the fermion. But firstly, let us consider the high-energy
   dependence of the lowest-order scattering amplitudes, which
   emphasizes a difference between renormalizable and
   non-renormalizable models.  Consider the diagrams (a) and (b) in
   Fig. \ref{one}, which describe scattering of $W$ on $F$.
   Conventional calculations show that at high collision energy their
   amplitudes satisfy
\begin{equation}
\label{M}
M^{(a)} \simeq -M^{(b)} \simeq
-\frac{e^2}{4m^2}\,(\,p_{\mu}+p'_{\mu}\,)\, {\bar F}\gamma^{\mu}F~.
   \end{equation} 
   Separately, each one of them grows with energy, violating the
   unitarity limit; note that the diagram (b) contains only one
   partial wave.  The increase is due to the longitudinal polarization
   of the W-boson, $\epsilon^{W}_{\mu}=k_{\mu}/m + O(m/p_0)$, $p_0$ is
   the $W$ energy - compare e.g.  \cite{cheng}.  The energy increase
   of the photon exchange diagram (a) signals the
   non-renormalizability of the pure vector electrodynamics.  However,
   the sum of the two diagrams $M^{(a)} +M^{(b)}$ does not possess
   this problem, a compensation of the two diagrams results in a
   reasonable behavior of the scattering amplitude at high energy, in
   accord with the renormalizability of the SU(2) model introduced in
   Eq.  \ref{triplet}.
   \noindent 
   It is important that the cancellation of the diagrams (a) and (b)
   manifests itself only when the collision energy is taken as a large
   parameter, $p_0 \gg m$.
    
   However, in the bound-state problem the energy is fixed, $p_0\simeq
   m$. Moreover, the wave function of the $W$-boson at distances
   $1/M\ll r \ll 1/ m$ is represented by the off shell diagrams, in
   which legs of the $W$-boson carry large momenta $m \ll |{\bf
     p}|,|{\bf p}'| \sim 1/r \ll M $.  The invariant $t$ of the
   scattering problem is large in this case, $t = -|{\bf p}-{\bf
     p}'|^2\gg m^2$, while the $s$-invariant remains fixed.  
   
   In this kinematic region the diagram (a) is {\em not} compensated
   by (b).  Therefore the verified above renormalizability of the
   theory cannot shed light on the behavior of the wave function at
   $1/M\ll r \ll 1/ m$.  To establish this behavior one needs to
   consider a set of all ladder diagrams of the type shown in Fig.
   \ref{one} (c), which are known to produce dominant contribution to
   the off-shell amplitude when $t$ is large. Clearly summation of
   this ladder is equivalent to the solution of the wave equation for
   the $W$-boson in an attractive Coulomb field created by the heavy
   fermion.
    
   The necessary wave equation can be derived from the Lagrangian
   Eq.(\ref{triplet}).  Keeping there only those terms, which describe
   $W$-bosons and their interaction with photons we find an effective
   Lagrangian, which proves to be identical to the Lagrangian
   introduces by Corben-Schwinger
\begin{eqnarray}
        \nonumber
        {\cal L}_W &=& -\frac{1}{2}
        \left(\nabla_\mu W_\nu - \nabla_\nu W_\mu\right)^+
        \left(\nabla^\mu W^\nu - \nabla^\nu W^\mu\right)
        \\ \label{cs}
         &&+i\,e \,F^{\mu\nu}W^{+}_\mu W_\nu +m^2 W_{\mu}^{+}W^\mu~,
   \end{eqnarray}
   Here $F^{\mu\nu}=\partial^\mu A^\nu-\partial^\nu A^\mu$ is the
   electromagnetic field, and $\nabla_\mu=\partial_\mu+i e A_\mu$ is
   the covariant derivative in this field ($e<0$). From Eq.(\ref{cs})
   one derives the Corben-Schwinger wave equation, which can be
   written in the following form
                \begin{eqnarray}
                \label{form} 
                \left( \nabla^2+m^2\right) W^\mu + 2 i e
                F^{\mu\nu}W_\nu +\frac{ie}{m^2} \nabla^\mu (j^\nu
                W_\nu) =0.\quad 
   \end{eqnarray} 
   The last term here includes explicitly an external current, $j^\nu=
   \partial_\mu F^{\mu\nu}$, which creates the electromagnetic field,
   see details in Ref. \cite{kuchiev-flambaum-06}.  To derive
   conclusions from Eq.(\ref{form}) several points need to be
   specified.  Since the boson is massive, it is necessary to rewrite
   the equation in terms of a $3$-vector, for example via spatial
   components ${\bf W}$ of the four-vector $W^\mu=(W^0,{\bf W})$. It
   is necessary also to specify the equation for a static, spherically
   symmetrical external potential $U(r)$, in which the $W$-boson
   propagates with angular momentum $j=0,1,\cdots$, these details can
   also be found in Ref.  \cite{kuchiev-flambaum-06}.
   
   Consider the most interesting for us partial wave $j=0$. In this
   state the three-vector ${\bf W}$ satisfies ${\bf W}({\bf r})={\bf
     n} \,v(r),~{\bf n}={\bf r}/r$. It can be verified that
   Eq.(\ref{form}) imposes the following equation on $v=v(r)$
   \begin{eqnarray}
\label{sec} v''+G \, v'+H \,v=0~, 
   \end{eqnarray} 
   where the coefficients $G=G(r)$ and $H=H(r)$ are
   \begin{subequations}\label{sub}
\begin{eqnarray} \label{g} G &=& \frac{2}{r}-\frac{U'}{U}-
\frac{U'+\Upsilon'} {U+\Upsilon}~, 
\\ \label{h} H &=&-\frac{2}{r^2}-\frac{2}{r}\,
\left(\frac{\,U'}{U}+\frac{U'+\Upsilon'}{U+\Upsilon}\right)
+(\,U+\Upsilon\,)\,U\,.\quad  
   \end{eqnarray}
   \end{subequations} 
   In the region $r\ll 1/m\simeq 1/p_0$ the energy and mass do not
   manifest themselves in the wave equation. The functions $G,H$ can
   depend only on the potential $U=U(r)$, in which the boson
   propagates, and on an additional term $\Upsilon=\Upsilon(r)$, which
   originates from the zero-th component of the external current
   $j_\mu$ in Eq.(\ref{form})
\begin{eqnarray}
        \label{ups}
        \Upsilon= e\,j_0/m^2=-\Delta U/m^2.
   \end{eqnarray}
   Let us apply firstly Eq.(\ref{sec}) to the pure Coulomb potential,
   when $U(r)=-Z\alpha/r$, where $Z=1/2$ is the charge of $F^+$, and
   $G=4/r$ and $H=(2+Z^2\alpha^2)/r^2$. Consequently we find the
   solution in the region $1/M\ll r\ll 1/m$,
\begin{eqnarray}
        v(r) \simeq r^{\gamma-3/2}~,   
        \label{gam}
   \end{eqnarray}
   where $\gamma=(1/4-Z^2 \alpha^2)^{1/2}$. Straightforward calculations
   show that this solution results in a major problem, forcing the
   charge density of the $W$-boson $\rho_W=\rho_W(r)$ to diverge at
   small distances, $\rho_W \propto r^{2\gamma-4}$.  Since $2\gamma<1$
   a divergence of the integral of this charge density signals the
   collapse of the $W$-boson to the Coulomb center, or at least into
   the region $r\sim 1/M$.  This makes the pure Coulomb problem poorly
   defined, in accord with conclusions of Ref.
   \cite{corben-schwinger_1940}.
   
   At this point it is instructive to return to the Compton-type
   diagram (b) in Fig.\ref{one}, which was so important in the high
   energy limit for on shell processes.  However, it is unable to
   remedy the problem of the collapse of the $W$-boson on the Coulomb
   center. The reason is clear.  We saw that the collapse takes place
   in the region $1/M\ll r\ll 1/m$, which is well separated from the
   heavy fermion, while the diagram (b) operates only when the
   distance $r$ between the $W$-boson and the heavy fermion $F^+$ is
   small, $r\sim 1/M$ \cite{nr}.
     Clearly the short-range interaction described by this diagram
   cannot prevent the fast increase of the wave function of the
   $W$-boson at larger distances $r > 1/M$.

   Consider now the radiative corrections. The most important
   phenomenon, which takes place at small distances (large momenta) is
   related to the renormalization of the coupling constant, which in
   the case considered results in the renormalization of the Coulomb
   charge.  It suffices to consider the vacuum polarization in the
   lowest-order approximation, when it is described by the known
   Uehling potential, which at small distances is represented via a
   conventional logarithmic function, see e. g. \cite{LL4}. A combined
   potential energy of the Coulomb and Uehling potentials read
\begin{eqnarray} \label{pot}
     U(r)=U_\mathrm{C}+U_\mathrm{U}=-[\,1-\alpha \beta \,\ln\left(m
       r\right)\,] Z\alpha/r~, 
   \end{eqnarray} 
   where $\beta$ is the lowest order coefficient of the Gell-Mann -
   Low beta-function. The polarization produces small variation of the
   potential in Eq.(\ref{pot}), but makes the $\Upsilon$-term
   Eq.(\ref{ups}) large at small distances
    \begin{eqnarray} \Upsilon= Z\alpha^2\beta/(m^2 r^3) \gg |U|~.  
      \label{Ylarge}
   \end{eqnarray}
   The functions $G,H$ in Eq.(\ref{sub}) calculated with account of
   this $\Upsilon$-term read, 
   \begin{eqnarray}
     \label{GH}
G\simeq 6/r~,\quad\quad
H\simeq  -Z^2\alpha^3\beta/(m^2r^4)~,      
   \end{eqnarray}
   which results in the following asymptotic solution of
   Eq.(\ref{sec})
\begin{equation}\label{sol}
v\propto \frac{1}{r^2}\times
\left\{ 
\begin{array}{l}
\exp (-\phi ),\quad \quad ~~\beta>0,  \\
\cos (\,|\phi|+\delta\,),       \quad \beta<0~,
\end{array} \right.
        \end{equation}
        where $\phi=Z\alpha (\alpha\beta)^{1/2}/(m\,r)$, and $\delta$
        is a constant phase defined by the behavior of the solution at
        $r\rightarrow0$, which we do not discuss here.  We see that
        the sign of $\beta$ plays a crucial role.  In pure QED it is
        positive, $\beta=2/(3\pi)>0$ for one generation of the Dirac
        fermions in the normalization adopted in Eq.(\ref{pot}).
        Eq.(\ref{sol}) indicates in this case that $v(r)$ is
        exponentially suppressed at small distances, which makes the
        Coulomb problem stable, well defined in accord with
        conclusions of Ref.  \cite{kuchiev-flambaum-06}.  In contrast,
        the considered SU(2) model is asymptotically free,
        $\beta=-22/(3\pi)<0$, which makes $v(r)$ a growing, strongly
        oscillating function at small distances. This clearly
        indicates the collapse of the $W$-boson. Therefore the Coulomb
        problem cannot be formulated in that case.
      
        We observe an unexpected result. For an attractive Uehling
        potential $U_\mathrm{U}<0$ (that characterizes the pure QED,
        $\beta>0$) the Coulomb problem turns out to be stable. In
        contrast, the repulsive Uehling potential $U_\mathrm{U}>0$
        (SU(2) model, $\beta<0$) results in the collapse of the
        $W$-boson, which makes the Coulomb problem unstable. In other
        words, the situation looks as if there is an {\it effective}
        potential, which sign is opposite to the sign of the Uehling
        potential.  This surprising behavior finds its origin in the
        properties of the $\Upsilon$-potential, which describes the
        zero-th component of the external current as shows Eq.
        (\ref{ups}). A presence of this current in the wave equation,
        see Eq.(\ref{form}), distinguishes the case of vector
        particles from scalars and spinors \cite{fnote}.
      
        The collapse of the $W$-boson on the Coulomb center is not
        related to particular properties of the model discussed.  It
        manifests itself similarly within, for example, the Standard
        Model $SU(2) \times U(1)$, if it includes heavy fermions
        \cite{heavy}.  At small distances $r <1/m$ the mass of the
        $Z$-boson may be neglected.  In this situation one may use any
        linear combinations of degenerate eigenstates. In our case it
        is convenient to use the original bosons $B_\mu^3$ from SU(2)
        and $W^{(0)} $ from U(1), instead of the eigenstates $Z$ and
        $\gamma$.  $W^-$ interacts with $W^0$ only, which reduces the
        problem to the SU(2) sector, where the $W$-boson collapses on
        the heavy fermion. We verified this claim by direct
        calculations, which show that at small distances the Weinberg
        mixing angle $\theta_W$ is canceled out. The Coulomb problem
        can be remedied only if a sufficient number of light fermions,
        which change the sign of the vacuum polarization, is added.
        The point is that this condition does not follow from the
        renormalizability of the theory.
        
        Finally, let us consider another aspect of the problem. Eqs.
        (\ref{sec}) - (\ref{ups}) contain $\Delta U_\mathrm{C} \propto
        \delta({\bf r})$, which was neglected in previous works.
        Working with this term it is convenient to introduce a finite
        nuclear size $R$ and then take the limit $R\rightarrow 0$ (for
        simplicity we assume infinite $M$).  Inside the ``nucleus'' an
        effective potential $U_\mathrm{eff}=-H$ given in Eq.(\ref{GH})
        dominates in the wave equation.  This attractive potential
        produces large number of states (infinite for the zero nuclear
        size), which are localized inside the nucleus.  Their energies
        are well below the ground-state energy given by the Sommerfeld
        formula.  These levels would be populated via creation of
        $W^+W^-$ pairs, similar to the vacuum breakdown for the Dirac
        particles. The difference is that for vector bosons the vacuum
        breakdown happens for any, however small charge of the
        nucleus.  Note that this contact potential was neglected when
        the Sommerfeld spectrum was derived for the point-like
        nucleus. However, we see that this potential drastically
        modifies the spectrum.
        
        In pure QED the problem is saved by the fermion vacuum
        polarization, which produces the impenetrable potential
        barrier (for $R=0$). This eliminates any contact interaction
        with the nucleus, and the Sommerfeld spectrum survives. In the
        case of the $SU(2)$ the situation may seem different since the
        Compton diagram Fig.\ref{one} (b) produces the repulsive
        interaction inside the nucleus. One may hope that this
        interaction eliminates the negative-energy states located
        inside the nucleus and brings the spectrum to the Sommerfeld
        form.  However, the contact interaction does not influence the
        wave function of $W$ at the distances $r \gg R$.  Therefore,
        the collapse of $W$ to a vicinity of the nucleus is inevitable.
        As was pointed above, the collapse can be prevented by
        addition of light fermions, which switch the ultraviolet
        behavior of the theory from the asymptotic freedom to the
        Landau pole, thus preventing the collapse.

      
        This work was supported by the Australian Research Council.
        One of us (VF) appreciates support from the Department of
        Energy, Office of Nuclear Physics, Contract No.
        W-31-109-ENG-38.  VF is also grateful to C. Roberts for a
        discussion of a possible link between this work and QCD
        \cite{comment}.

    \appendix


\begin{thebibliography}{99}
\bibitem{nr}
  
  In the non-relativistic limit the Compton diagram gives a repulsive
  contact interaction $\sim (|e|/p_0)(W_0^{\dagger} W_0)\rho(r)$ where
  $\rho=j_0=Z |e| \delta^{3}({\bf r})$ is the charge density of the
  Coulomb center; the energy of the $W$-boson $p_0\approx m$.
\bibitem{fnote} 
   
   It is interesting to observe the following trend.  The wave
   equation for scalars, $S=0$, includes only the potential $A_\mu$.
   The wave equation for spinors, $S=1/2$, if written as the
   second-order differential equation, includes also the field
   $F_{\mu\nu}$, which is represented by first derivatives of the
   potential. In the wave equation for vector bosons, $S=1$, there
   appears also the external current $j_\mu$, which can be expressed
   via second derivatives of the potential.
\bibitem{heavy} 
   
   The inclusion of new heavy particles is motivated by the dark
   matter observations and possible extensions of the Standard model,
   e.g. supersymmetry.
\bibitem{comment} 
  
  Zero-mass particles have no bound states with a Coulomb center since
  the binding energy is proportional to the mass.  The spontaneous
  symmetry breaking, which creates the mass via the Higgs mechanism
  generates the bound states.  The gluon acquires a dynamical mass
  inside a hadron and, in principle, may form a bound state with a
  heavy quark. This problem may also be relevant to the quark-gluon
  plasma.
\end{thebibliography}
\end{document}